\RequirePackage{fix-cm}

\documentclass[twocolumn]{svjour3}          
\smartqed  
\usepackage{graphicx}
\usepackage[utf8]{inputenc}
\usepackage[T1]{fontenc}
\usepackage{mathptmx}
\usepackage{amsmath}
\usepackage{xcolor}
\usepackage{url}
\begin{document}

\title{Microwave Tomography with phaseless data on the calcaneus by means of artificial neural networks}

\author{J. E. Fajardo \and F. P. Lotto \and F. Vericat \and C. M. Carlevaro \and R. M. Irastorza}



\institute{J. E. Fajardo, F. Lotto, F. Vericat, C. M. Carlevaro, and R. M. Irastorza \at
              Instituto de F\'isica de L\'iquidos y Sistemas Biol\'ogicos CONICET - CCT La Plata, Argentina\\
%
           C. M. Carlevaro \at
	   Depertamento de Ingeniería Mecánica, UTN - FRLP, Berisso, Argentina.\\
%
           *R. M. Irastorza \at
           Instituto de Ingenier\'ia y Agronom\'ia, UNAJ. Florencio Varela, Buenos Aires, Argentina\\
              \email{rirastorza@iflysib.unlp.edu.ar}   
}

\date{Received: date / Accepted: date}

\maketitle

\begin{abstract}

The aim of this study is to use a Multilayer Perceptron (MLP) Artificial Neural Network (ANN) for phaseless imaging the human heel (modeled as a bilayer dielectric media: bone and surrounding tissue) and the calcaneus cross-section size and location using a two dimensional (2D) microwave tomographic array. Computer simulations were performed over 2D dielectric maps inspired by Computed Tomography (CT) images of human heels for training and testing the MLP. A morphometric analysis was performed to account for the scatterer shape influence on the results. A robustness analysis was also conducted in order to study the MLP performance in noisy conditions. The standard deviations of the relative percentage errors on estimating the dielectric properties of the calcaneus bone were relatively high. Regarding the calcaneus surrounding tissue, the dielectric parameters estimations are better, with relative percentage error standard deviations up to $\approx$ 15\%. The location and size of the calcaneus are always properly estimated with absolute error standard deviations up to $\approx $ 3 mm. 
\end{abstract}

\section{Introduction}
\label{intro}
The dielectric properties (relative permittivity ($\varepsilon_r$) and conductivity ($\sigma$)) of the cancellous bone are related to its microstructure, organic and inorganic composition which togheter determine the bone quality \cite{meaney2012bone,sierpowska2007electrical,irastorza14,amin2019dielectric}. \\
Microwave Tomography (MWT) is a consolidated technique suitable for retrieving the electromagnetic parameters ($\varepsilon_r$ and $\sigma$) and shape of an unknown target located in an investigation domain \cite{LibroPastornino10}. Recently, Meaney et al. \cite{meaney2012clinical} have successfully used MWT technique to measure dielectric properties of human bone \textit{in vivo}. They have studied the calcaneus, which is the main bone of the heel, by measuring amplitude and phase of the transmitted wavefield of a circular array of antennas. Phase measurement presents considerable practical difficulties and hardware cost which makes amplitude--only methods highly attractive \cite{li2008,li2009,costanzo2015}. Simulating the same array of antennas, Fajardo et al. \cite{fajardo} have computationally studied the feasibility of detecting dielectric properties of the calcaneus with phaseless data of two dimensional (2D) models. The authors have concluded that the problem is highly sensitive to the conductivity of the tissues that surround the calcaneus and the conductivity of the calcaneus itself. Moreover, the work also showed evidence supporting the fact that the cortical bone layer (cortex of the calcaneus) and skin (surrounding the heel) do not weight in the detection of the dielectric properties of the calcaneus.\\
Usually, the inverse problem in MWT can be solved applying two different approaches: deterministic (e.g.: the contrast source inversion method \cite{li2008}) or stochastic (e.g.: particle swarm optimization \cite{franceschini2006inversion}). An alternative approach is to make use of Artificial Neural Networks (ANN). In reference \cite{goodfellow2016deep} it has been shown the capability of the Multilayer Pereceptron (MLP) ANN for estimating parameters of complex nonlinear problems. Bermani et al. \cite{bermani2002microwave} have achieved good results measuring the location and dielectric properties of a homogeneous cylinder with amplitude-only information by means of a MLP ANN. Recently, Wei and Chen have obtained a fast method for inverse scattering problems using convolutional neural networks \cite{wei2018deep}. Li et al. \cite{li2018nonlinear} have proposed a general approach for solving nonlinear electromagnetic inverse scattering using convolutional neural networks aswell.\\
The objective of this work was to present an efficient inversion method for the detection of dielectric properties of the calcaneus with phaseless data. We propose a MLP ANN using a deeper network than that applyed by \cite{bermani2002microwave} with more training data. The 2D model considers the whole heel as a two dielectric media (calcaneus and surrounding tissue) with realistic shape inspired by Computed Tomography (CT) images of human heels. In order to quantify the shape differences between the 2D dielectric maps, morphological variability of the images was assessed using geometric morphometric techniques \cite{adams2004geometric}.

\section{Materials and Methods}
\subsection{Data Generation}\label{simm} 
\subsubsection{Electromagnetic Simulations}
\begin{figure*}[!t]
\centering
\includegraphics[width=1\textwidth]{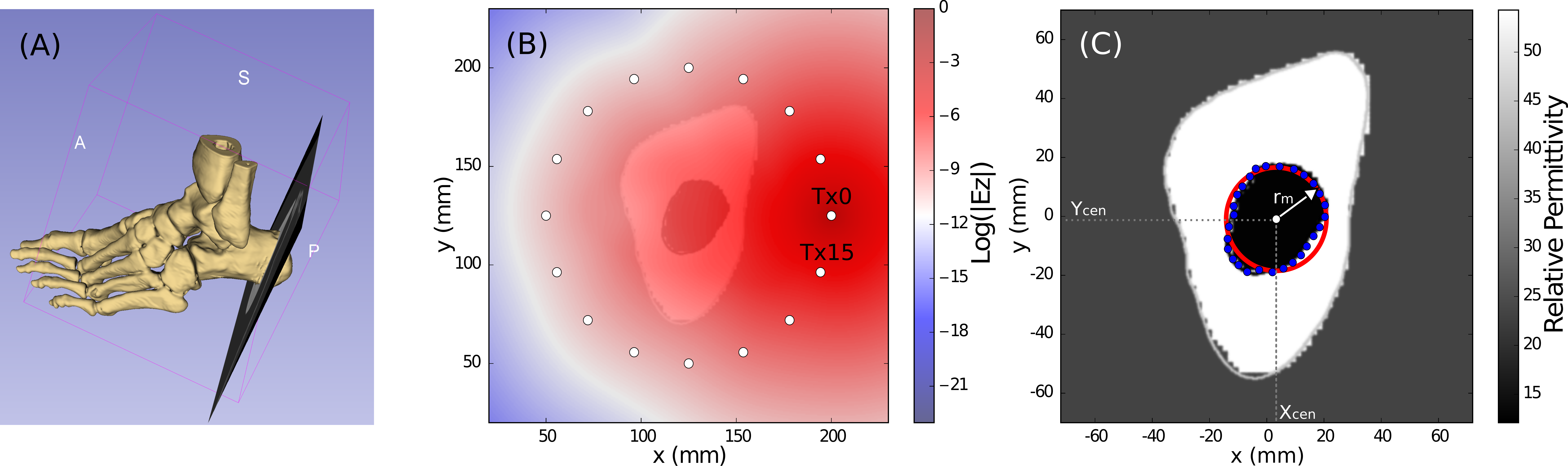}
\caption{\footnotesize (A) 3D Reconstruction of the feet from CT, showing one of the planes used for reconstructing the heels 2D images. (B) Simulation showing the antenna Tx = 0 emitting at the right side of the image and the rest of the antennas (white dots). 
The colored bar corresponds to the magnitude of the electric field ($\left| E_{z} \right|$) in logarithmic scale for better visualization. Tx = 15 is also shown for reference. (C) Coordinates sampling the bone edges (blue dots), the centroid (white dot with coordinates $\textrm{X}_{\textrm{cen}}$, $\textrm{Y}_{\textrm{cen}}$) and equivalent radius (r$_{\textrm{m}}$) of the circular cross section area of the bone (red circle). The grayscale represents the relative permittivity values of this particular 2D slice.}\label{sim}
\end{figure*}

We conducted several simulations in order to train the ANN. A total of 18 coronal CT slices of human heels (from 9 different patients) at different heights and inclinations were used as the realistic shape geometries for simulating the electromagnetic phenomena (see Fig. \ref{sim} (A)). Segmentations and image processing was performed using 3DSlicer \cite{fedorov20123d}.
The geometry of each model considers the whole heel as a two dielectric media: (I) surrounding tissue (muscle/tendon), and (II) calcaneus (considered as trabecular bone). This simplified model is based on the results presented in the sensitivity analysis of our previous work \cite{fajardo}. The range of values of conductivity and permittivity of the materials employed in computer model are shown in Table \ref{dielprop} \cite{fajardo}.\\
\begin{table}[h!]
\centering
\caption{\footnotesize Simulated range of the dielectric properties of the tissues involved (at 1.3 GHz).}
\label{dielprop}
\begin{tabular}{lccc} 
\hline
Region & $\varepsilon_{r}$ & $\sigma$ (Sm$^{-1}$) \\
\hline
I (Muscle/Tendon) & [29.0, 70.0] & [0.90, 1.55] \\
II (Trabecular Bone) & [12.5, 20.1] & [0.44, 0.92]\\

\hline
\end{tabular}
\end{table}
Maxwell equations were numerically solved in 2D using finite-difference time-domain (FDTD) method (implemented in the freely available software package MEEP \cite{paperMeep10}). An array of monopole antennas similar to that developed by Meaney et al.\cite{meaney2012bone} was simulated (a circular array of 16 antennas equally angularly spaced and disposed in a circle with a diameter of 152 mm). The monopole antennas were modeled in 2D as a line of current (a point source in 2D) emitting a TM-polarized electric field $E_z \propto e^{j\omega t}$, being ``$z$'' the axis parallel to the antennas, where $\omega=2\pi f$, and $f$ is the frequency ($f=$ 1.3 GHz). The size of the simulation box was 250 mm $\times$ 250 mm, the spatial grid resolution of the simulation box was 1.0 mm and the Courant factor was 0.5. The boundary conditions were Perfectly Matched Layers (PML), which means that there was total absorption in the box edges. The coupling bath was a glycerin-water mixture in 80:20 proportion.\\
In the simulations we computed the absolute value of the total field ($\left| E_{z} \right|$) in the receiver antennas (Rx). This magnitude was measured in the seven receiver antennas right in front of the transmitter (Tx), since in \cite{fajardo} was shown that these antennas have considerably more information than the others. A complete simulation gives a total of 7$\times$16 = 112 measurement points. In Fig. \ref{sim} (B), the radiation pattern created by Tx = 0 (transmitter number 0) is shown over the simulation box.\\
A total of 7200 complete simulations were performed from 18 different 2D slices (for training and testing purposes) containing different combinations of the dielectric properties taken from a random uniform sampling of the intervals shown in Table \ref{dielprop}. Two hundred different combinations of these parameters were generated from each 2D slice. The heel position was also displaced within the tomographic circular array. Furthermore, we duplicated the amount of models by mirroring the original images, taking advantage of the geometric symmetry.\\
In order to train the ANN in a robust way and to evaluate the prediction accuracy in presence of different noise levels, two main sources of errors were added: additive noise in the receptor and unknowledge in the actual position of it. Therefore, the data from each simulation were altered as follows. The input vector $j$ of the MLP is:
\begin{equation}
\mathbf{x}_j^{noisy}={\left| E_{z} (x,y)\right|}_j+\mathbf{N}(0,\text{std}_2^2)
\label{noise}
\end{equation}
where the normally distributed random variables $x$ and $y$ (with mean in the true antennas coordinates and standard deviation std$_{1}$) emulate the uncertainty in the position of the antennas. $\mathbf{N}\left(0,\text{std}_2^2\right)$ is a normally distributed variable with zero mean and standard deviation std$_{2}$ corresponding to the noise in the measured signal. Different noise levels were added, std$_{1}$ up to 1 mm, and std$_{2}$ up to 5\% of the unaltered received signal (it approximately corresponds to a signal--to--noise ratio (SNR) up to 26 dB). The main results of this work were obtained using these noise levels. In the following, it will be explicitly indicated if they are changed for a particular parameter study.

\subsubsection{Morphometric Information} \label{geo} 

In order to quantify shape differences between the 2D dielectric maps, morphological variability of the outlines was assessed using geometric morphometrics techniques \cite{adams2004geometric}. Additionally, a set of 28 $(x,y)$ coordinates on the bone borders was recorded to compute the $\textrm{X}_{\textrm{cen}}$ and $\textrm{Y}_{\textrm{cen}}$. The computed coordinates of the centroid of the calcaneus, and an equivalent radius ($\textrm{r}_{\textrm{m}}$), are shown in Fig. \ref{sim} (C).\\

\begin{figure}[!t]
\centering
\includegraphics[width=1\columnwidth]{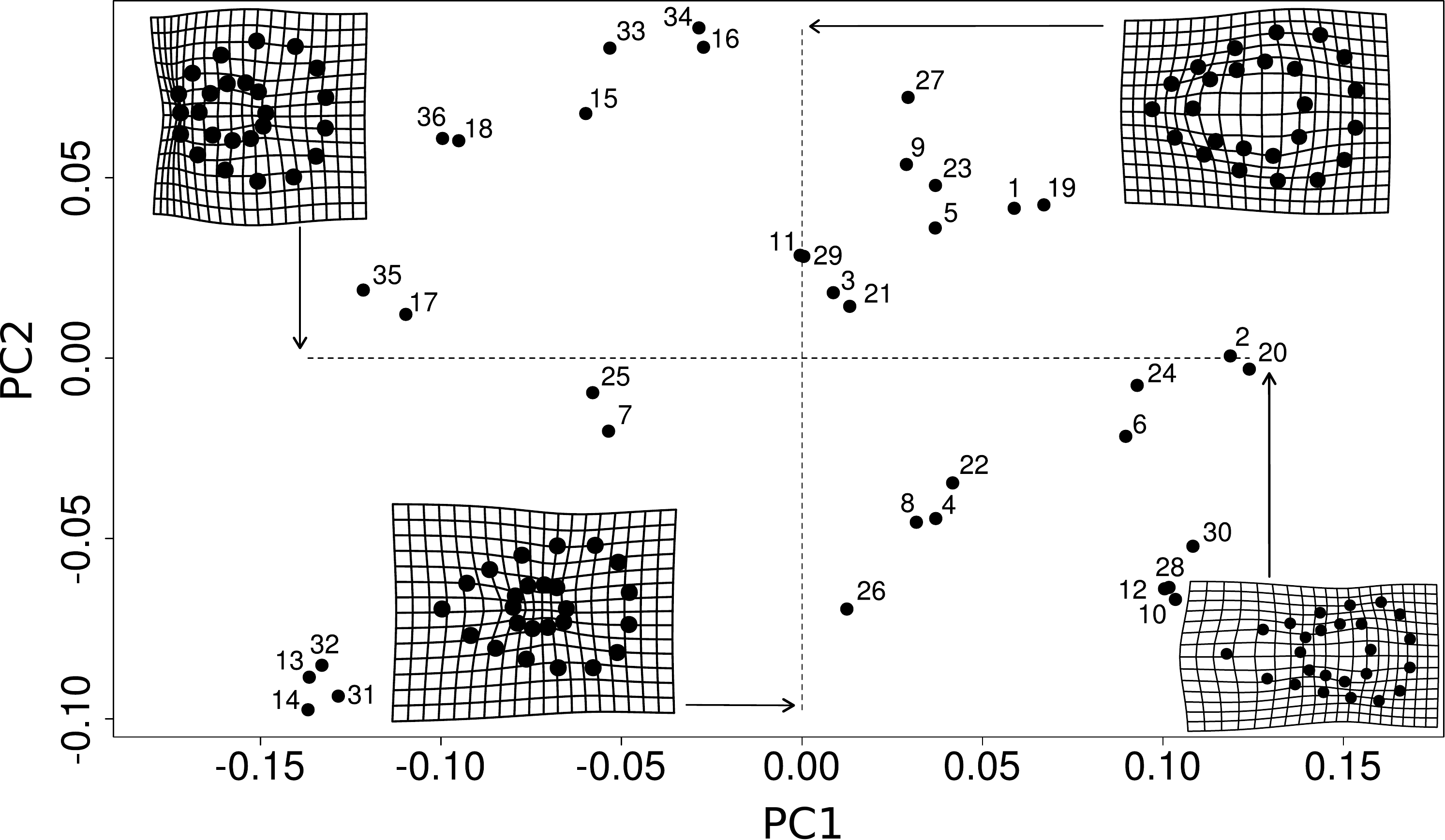}
\caption{\footnotesize Geometric distribution of the 2D slices in the space of the PC 1 and 2, The number of the slice is indicated next to
the corresponding point. Also as reference, deformed grids in the extremes of PC1 and PC2 are shown with the landmarks over them, 
ilustrating the deformation respect to the mean form in the corresponding axis.}\label{pca}
\end{figure}

For the morphometric analysis, a set of 25 cartesian coordinates in 2D were recorded for each original slice, 15 on the
skin outline and 10 on the bone outline. Outlines were digitized as series of discrete equidistant points, 
known as semilandmarks. The individual points were slid along a tangential direction so as to remove tangential variation, 
because contours should be homologous from slice to slice, whereas their individual points need not. This variation can be removed
by minimizing procrustes distance with respect to a mean reference form \cite{bookstein1997landmark}.
Semilandmarks digitization on the skin outline started from an anatomical landmark, defined as the
posterior outermost point of the Achilles tendon, and proceeded clockwise. On the bone outline, a second
landmark was defined as the intersection of the bone outline and a vertical line (with respect to the image
frame of reference) passing through the first landmark. Semilandmarks on the bone outline were then
digitized clockwise from this second point. The coordinates of landmarks were aligned using generalized
least-squares procrustes fit \cite{bookstein1997morphometric}. This procedure optimally translates, rotates and scales
coordinates of landmarks in order to remove the information on position, orientation and size \cite{rohlf1990extensions}. 
A Principal Components (PC) analysis was then conducted on the covariance matrix of the
procrustes residuals, and the two first PC (covering $\approx 70\%$ of the variance) were used to quantify shape differences
between slices.\\
Figure \ref{pca} shows the position of each slice in the two first PC shape space. It also shows as reference the deformed grids with the aforementioned landmarks at the extreme values of the two first PC axes.

\subsection{The MLP ANN}\label{ann} 
\subsubsection{MLP ANN Overview}
\label{annov} 
A MLP ANN is an interconnected network of artificial neuron-like units, disposed in layers, where all the units of each layer are connected with all the units of the previous and next layer, but there are not connections between the units in the same layer. \\
In this kind of ANN the output of each neuron variates according to a given input $\mathbf{x}_j=\{ x_{1j}\cdots x_{Nj}\}$, applied to a network of $N_m$ layers and $N_n$ nodes in the $n$-th layer as:
\begin{equation}
x_k(l)=\mathbf{f}\left(\sum_{i=1}^{N_{l-1}}x_i(l-1)w_i^{(k)}(l)+b_k(l)\right)
\label{mlp}
\end{equation}
with $l=1\cdots N_m$, $k=1\cdots N_l$, being $\mathbf{f}\left(\cdot \right)$ the activation function, $x_k(l)$ the state of the $k$-th neuron of the $l$-th layer, $w_i^{(k)}(l)$ the associated weight to the connection between the $i$-th neuron of the $(l-1)$ layer and the $k$-th neuron of the layer $l$ and $b_k(l)$ the bias associated to the $k$-th neuron of the $l$-th layer.\\
Once input vectors $\mathbf{x}_j$ are propagated through the network and output vectors $\tilde{\mathbf{y}_j}$ of parameters are calculated, the result is compared to the actual $\mathbf{y}_j$ values and a loss is calculated. Tipically one of the loss metrics used in regression is the Mean Absolute Percentage Error (MAPE):
\begin{equation}
\text{MAPE}=\frac{1}{n}\sum_{i=1}^n\frac{\left|\tilde{y_i}-y_i\right|}{y_i}
\label{mae}
\end{equation}
where $\tilde{y_i}$ and $y_i$ are, respectively, the output of the model (a particular parameter) and the actual value, $n$ is the number of parameters where the difference is evaluated. In this work we also use the Mean Absolute Error (MAE) which is directly computed as $\frac{1}{n}\sum_{i=1}^n\left|\tilde{y_i}-y_i\right|$.\\

After evaluating the loss, the gradients are retropropagated through the network from $\mathbf{y}$ to  $\mathbf{x}$ and the weights are updated in all the neuron-like units composing the network
in a direction opposite to that of the gradients $\nabla_{\theta}J(\theta)$, where $J(\theta)$ is the objective or loss function parameterized to the model parameters $\theta$. 
Several approaches are used for doing this task. For an overview in this particular topic, see \cite{GDoverview}.\\
Once the loss is low enough and the network is capable of generalizing the prediction for new input vectors (not used during the training period), the matrices with the weights are saved and a useful model is obtained.

\subsubsection{Topology and characteristics of the implemented MLP ANN}\label{ourann} 

The topology of the implemented MLP ANN corresponds to a first layer of inputs, which are the antennas measurements described in the subsection \ref{simm} (i.e.: 112 inputs regarding $\left| E_{z} \right|$), then three ``hidden'' layers with 100 units each one, and finally an output layer with seven units (corresponding to the parameters  to be estimated: $\varepsilon_r$ and $\sigma$ of the calcaneus and the sourrounding tissue, $\textrm{X}_{\textrm{cen}}$ and $\textrm{Y}_{\textrm{cen}}$ coordinates of the calcaneus centroid, and the equivalent radius ($\textrm{r}_{\textrm{m}}$)). As activation functions, rectifier linear units functions were used in all the neurons, except in the output layer, where a general linear function was used. The rectifier linear unit function is a function $\mathbf{f}\left(\cdot\right)$ whose value is 0 if $\mathbf{f}\le0$ and a linear function if $\mathbf{f}>0$. The weights updating ($w_i^{(k)}$) during the network training was implemented using backpropagation with the ``Adam'' gradient descent method \cite{adam}. This approach outperformed other gradient descent methods in this particular study.\\
We used the application programming interface Keras \cite{keras} with Tensorflow \cite{tensorflow} as backend package for implementing the ANN. The selected loss function was MAPE (Eq. \ref{mae}). The data were normalized to the $\left[0,1\right]$ range and 20 epochs were used for training the network with a batch size of 50. The fixed number of epochs was previously found using early--stopping and then fixed to use validation data as test data avoiding validation bias, despite the risk of underfitting in some cases.\\
The network training and testing was done using jackknife-like cross validation \cite{mosteller1971jackknife} in order to use the higher ammount of models in the training set and also to use all the data as test set. For example: taking out the data corresponding to $X_i$ (with $i=1,\ldots, n$) for testing the model from the training set $\mathbf{X} = \{ X_1,\ldots,X_{i-1}, X_{i+1},\ldots,X_n \}$ (with $X_i \notin \mathbf{X}$), being $X_i$ a matrix with all the $\mathbf{x}_j$ input vectors of the ANN made of combinations of dielectric properties generated from the $i$-th 2D slice.

\section{Results}
\begin{figure*}[!t]
\centering
\includegraphics[width=1\textwidth]{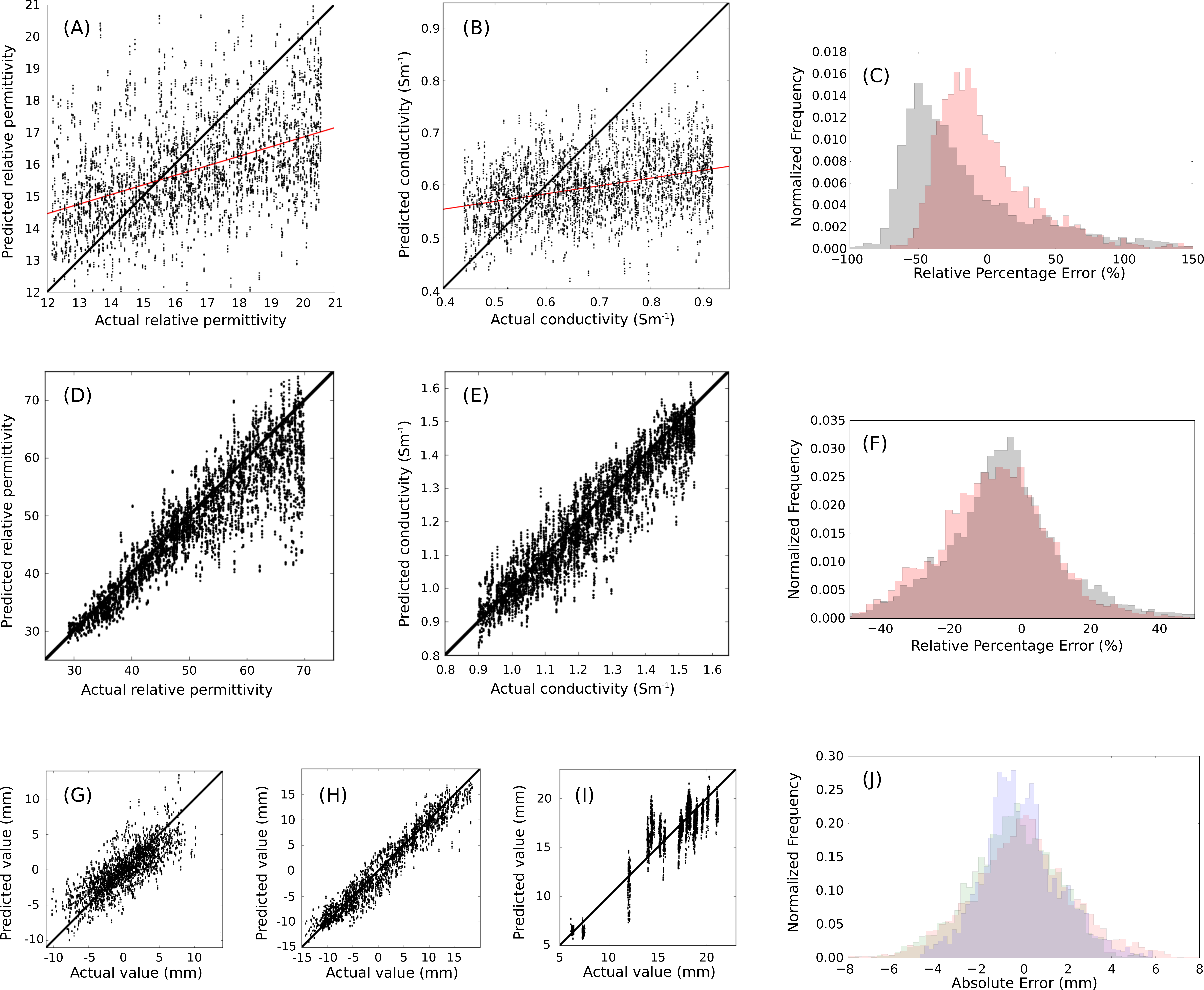}
\caption{\footnotesize Predictions for the 7200 models. (A) and (B) predictions and true values of the relative permittivity and conductivity, respectively, of the calcaneus. A red line is plotted with the best linear fit. (C) Histogram of the relative errors for the prediction of the calcaneus conductivity (grey) and relative permittivity (light red). (D) and (E) predictions and true values of the relative permittivity and conductivity, respectively, of the calcaneus surrounding tissue. (F) Histogram of the relative errors of the prediction of the calcaneus surrounding tissue conductivity (grey) and relative permittivity (light red). (G), (H), and (I) predictions and true values of the $\textrm{X}_{\textrm{cen}}$ coordinate, $\textrm{Y}_{\textrm{cen}}$ coordinate, and equivalent radius $\textrm{r}_{\textrm{m}}$ of the calcaneus. (J) Histogram of the prediction error of the $\textrm{X}_{\textrm{cen}}$ (light green), $\textrm{Y}_{\textrm{cen}}$ (light red), and $\textrm{r}_{\textrm{m}}$ (light blue).}\label{pred}
\end{figure*}

\begin{figure}[!t]
\centering
\includegraphics[width=1\columnwidth]{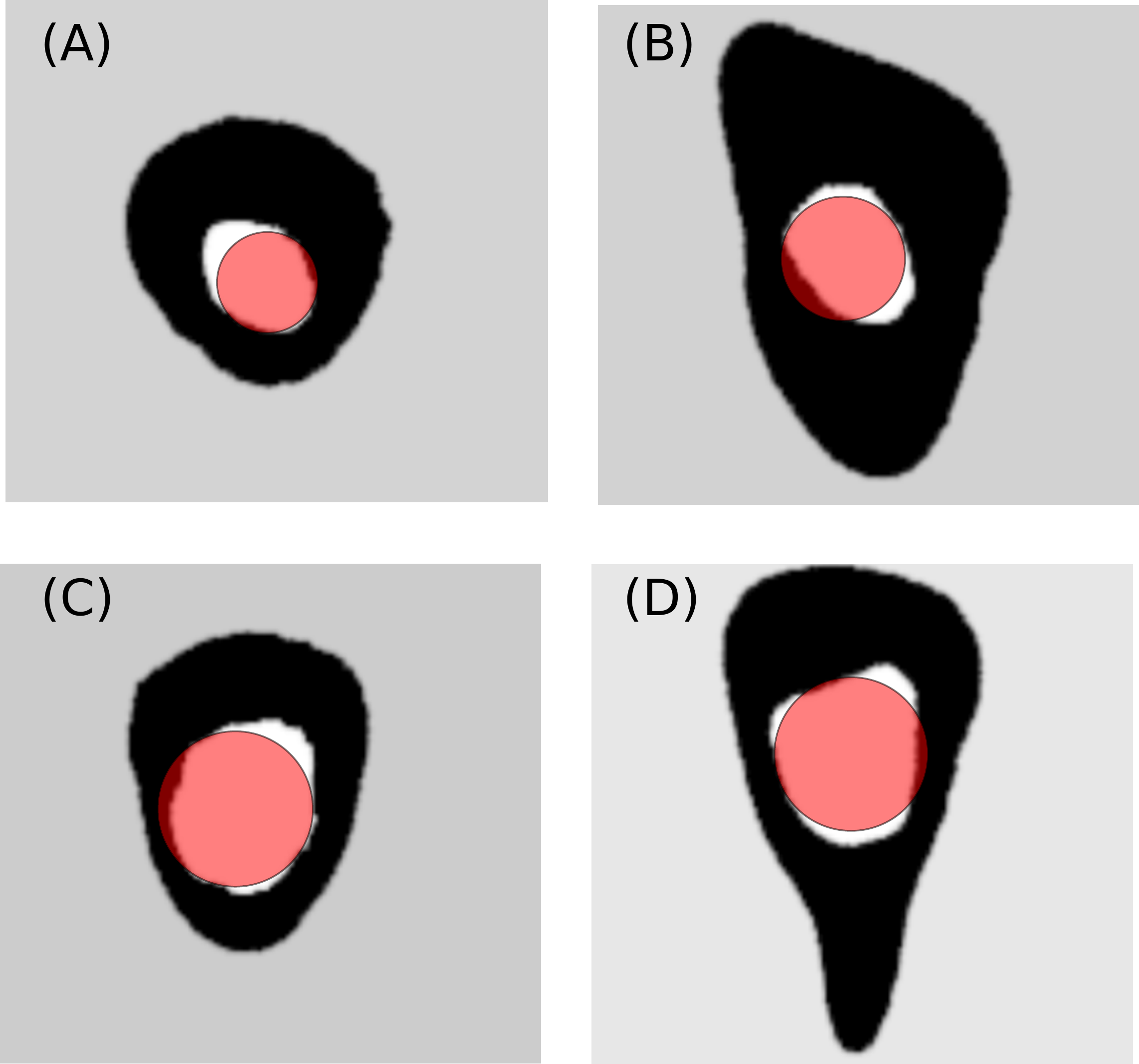}
\caption{\footnotesize Estimation of the geometric properties. The red circles plotted over the actual simulated heels are obtained from the ANN predictions (parameters: $\textrm{X}_{\textrm{cen}}$, $\textrm{Y}_{\textrm{cen}}$, and $\textrm{r}_{\textrm{m}}$).}\label{rec}
\end{figure}

Seven output parameters were obtained from the inversion method presented in this work, four related to the dielectric properties and three to the calcaneous cross--section. We will call them dielectric and geometric parameters, respectively. Figure \ref{pred} shows the predictions of these parameters for the 7200 models generated from the 18$\times$2 slices using the metodology previously described. Linear fitting was computed for the dielectric properties of the calcaneus (see Fig. \ref{pred} (A) and (B)). This shows that there were always an overestimation of the actual lower values and a sub estimation of the higher ones. The histograms of the errors for each parameter estimation are also shown. While the errors in the estimation of the dielectric properties of the surrounding tissues (Fig. \ref{pred} (F)) and the geometric parameters (Fig. \ref{pred} (J)) seem to follow a normal distribution, this is not the case for the estimation errors of the dielectric properties of the calcaneus (Fig. \ref{pred} (C)).\\
An example of the estimation of the geometric parameters of the calcaneus (centroid and equivalent radius) is shown in Fig. \ref{rec}. Estimations from four different slices are shown.\\

\begin{figure}[!t]
\centering
\includegraphics[width=0.9\columnwidth]{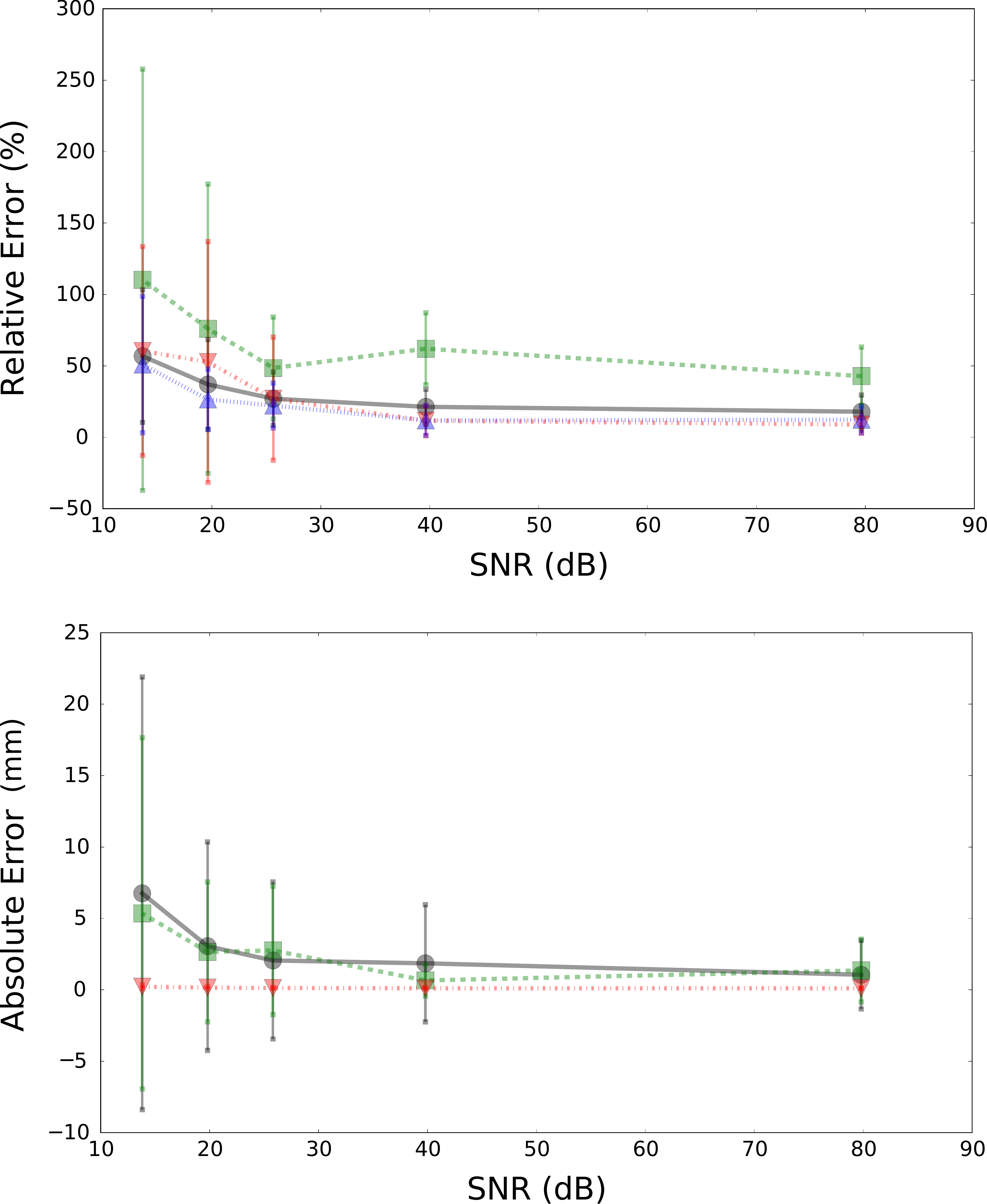}
\caption{\footnotesize The mean (marks) and standard deviation (error bars) of the prediction erros for 200 different combinations of the parameters generated from slice 21. (A) MAPE of the calcaneus $\varepsilon_r$ (grey circles) and $\sigma$ (green squares) and surrounding tissue $\varepsilon_r$ (blue triangles) and $\sigma$ (red triangles) for different SNR in the signal. (B) MAE of the calcaneus geometric parameters $\textrm{X}_{\textrm{cen}}$ (grey circles), $\textrm{Y}_{\textrm{cen}}$ (green squares) and $\textrm{r}_{\textrm{m}}$ (red triangles) for different SNR in the signal.}\label{err}
\end{figure}

In Fig. \ref{err}, the effect of the SNR on the prediction error is shown. The 200 models generated from a single slice (named 21) were depicted. This slice was selected under the assumption that it is close to the mean shape of the PC space (see Fig. \ref{pca}) and the relative errors in the predictions were representative of the overall behavior of the MLP ANN. The training set for this particular study was generated without noise in the received signals. As expected, the errors decrease when the SNR become higher. Figure \ref{err} also shows that the conductivity of the calcaneus presents the greatest errors in estimation even for high SNR values.\\

The prediction error behaviour related to differences in the shape of the slices is studied by computing the linear correlations (r) between the scores of the two first PC (PC1 and PC2), and the mean and the standard deviation of MAPE (for $\varepsilon_r$ and $\sigma$) and MAE (for $\textrm{X}_{\textrm{cen}}$, $\textrm{Y}_{\textrm{cen}}$, and $\textrm{r}_{\textrm{m}}$). Tables \ref{tabcorr1} and \ref{tabcorr2} show the results (see Fig. \ref{pca} for reference). The highest linear negative correlation is obtained between PC2 and the standard deviation of the error in the estimation of $\varepsilon_r$ of the calcaneus (significantly $p<0.05$). Regarding the geometric properties, the same behavior is observed for $\textrm{X}_{\textrm{cen}}$ and $\textrm{Y}_{\textrm{cen}}$. PC1 is weakly positive correlated with the mean error in the estimation of $\varepsilon_r$ and $\textrm{X}_{\textrm{cen}}$ of the calcaneus.
\begin{table*}[h!]
\centering
\caption{\footnotesize Linear correlations (r) between PC1 and PC2, and the mean and standard deviation of MAPE for the predictions of the dielectric properties of the calcaneus and the surrounding tissue. The correlations with determination coefficient greater than 25\% are highlighted with boldface font.}
\label{tabcorr1}
\begin{tabular}{lcccc} 
\hline
 & \multicolumn{2}{c}{Calcaneus} & \multicolumn{2}{c}{Surrounding tissue}\\
  & Mean MAPE $\varepsilon_r$ (Std.) & Mean MAPE $\sigma$ (Std.)& Mean MAPE $\varepsilon_r$ (Std.) & Mean MAPE $\sigma$ (Std.) \\
\hline
PC1 & \textbf{0.52}** (0.12) & 0.03 (0.13) & -0.17 (-0.18) & 0.00 (0.25) \\
PC2 & -0.11 (\textbf{-0.71}**) & 0.06 (-0.25) & \textbf{0.67}** (0.28*) & -0.10 (0.36**) \\

\hline

* $p<$ 0.10.\\
** $p<$ 0.05.
\end{tabular}
\end{table*}

\begin{figure}[!t]
\centering
\includegraphics[width=0.9\columnwidth]{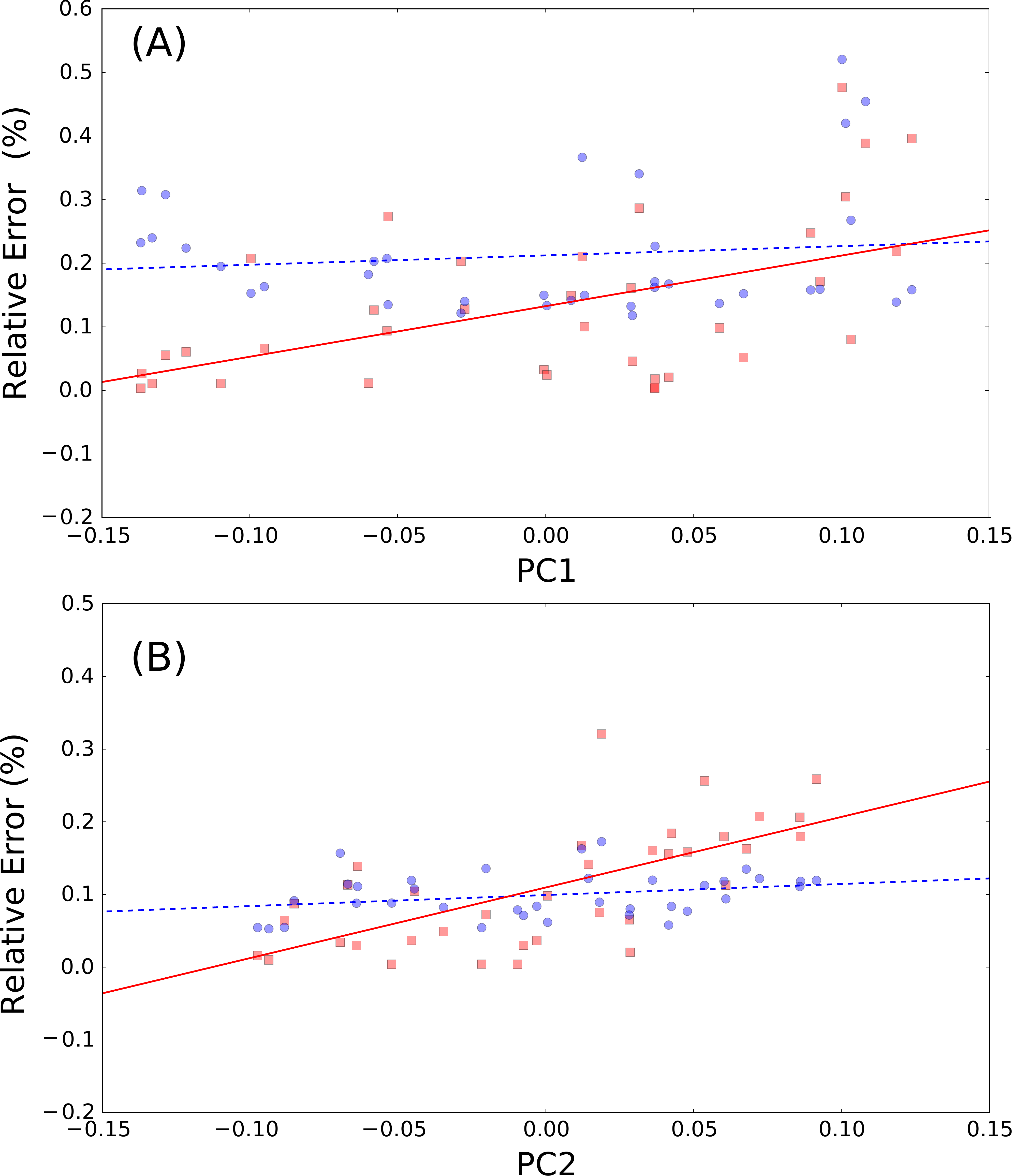}
\caption{\footnotesize Relationship between PC1 and PC2, and the mean (red squares) and standard deviation (blue circles) of the MAPE for the relative permittivities of the calcaneus (A) and the surrounding tissue (B). Each point represents the mean (red square) and standard deviation (blue circle) of the prediction relative errors of 200 models generated with different dielectric properties and position of a particular slice.}\label{epspc}
\end{figure}

\begin{table*}[h!]
\centering
\caption{\footnotesize Linear correlations (r) between PC1 and PC2, and the mean and standard deviation of MAE for the predictions of the geometric properties of the calcaneus. The correlations with determination coefficient greater than 25\% are highlighted with boldface font.}
\label{tabcorr2}
\begin{tabular}{lccc} 
\hline
   & Mean MAE $\textrm{X}_{\textrm{cen}}$ (Std.) & Mean MAE $\textrm{Y}_{\textrm{cen}}$ (Std.)& Mean MAE $\textrm{r}_{\textrm{m}}$ (Std.)\\
\hline
PC1 & \textbf{0.56}** (0.17) & 0.37** (0.17) & 0.32* (0.10)\\
PC2 & -0.34** (\textbf{-0.67}**) & -0.28* (\textbf{-0.67}**) & -0.14 (0.42**) \\

\hline

* $p<$ 0.10.\\
** $p<$ 0.05.
\end{tabular}
\end{table*}
Figure \ref{epspc} shows the linear correlations between the first and second PC and the mean and standard deviation of the relative permittivity MAPE for the calcaneus and the surrounding tissue. Each point represents the mean and standard deviation of the MAPE of 200 models generated with different dielectric properties and position of a particular slice.\\

\section{Discussion}
The motivation of this study is to search an alternative method to evaluate bone health \textit{in vivo}. Microwave imaging is a technique that could provide new insights to this exciting clinical challenge. Mainly because it can be applied periodically in patients due to its non ionizing condition. Other desired conditions in such alternative methods are: low cost and portability of the required equipment. This work presents a robust inversion method for MWT with phaseless data that gives a step towards meeting the aforementioned characteristics.\\
The magnitude of the errors obtained in the prediction of the relative permittivity and geometric circular approximation of the calcaneus suggests the suitability of using a MLP ANN for estimating these properties with an experimental setup similar to that used for Meaney et al. \cite{meaney2012clinical}. However, the high error in predicting the calcaneus dielectric conductivity makes this method not--suitable for predicting this particular parameter properly. The dielectric properties of the calcaneus surrounding tissue are predicted but, in this case, the conductivity prediction errors are lower than those of the permittivity.\\
The model is robust enough to noise levels similar to those used during its training (about 26 dB in the signal and 1 mm in the antennas position uncertainty). Increasing the SNR over these levels, leads to a faster decreasement of the dielectric properties prediction errors.\\
Special attention was paid in understanding the behavior of the ANN with the geometry and the position of the heel. We studied the results yielded only by the first two Principal Components. It can be assumed from Fig. \ref{pca} that PC1 and PC2 are associated to the inclination of the heel and the area occuped by the calcaneus within the heel, respectively. The PC1 and relative permittivity of the calcaneus are weakly correlated. It can be seen that for lower values of PC1 (the slice of heel is less stretched) lower errors are obtained. In general, the dispersions in the predictions of the calcaneus parameters are better at higher values of PC2 (the correlation coefficients were always negative, excepting for the equivalent radius). This means that the higher the area occupied by the calcaneus the lower the dispersions of the estimate errors. This result should be carefully considered because the estimate could be biased but with low variance. Low values of PC2 improved the estimation of the relative permittivity of the surrounding tissue (see Fig.\ref{epspc} (B)). This is expected, since lower PC2 values are associated with lower calcaneus cross--section areas. In short, better estimations of the relative permittivity could be achieved avoiding inclination towards the Achilles tendon (in the case of calcaneus), and having a smaller proportion of bone area (in the case of the surrounding tissue).\\
A commentary should be added regarding the inversion method. The convergence of traditional or stochastic algorithms usually is time consuming and depends on the initial guesses. In the method proposed here, once the ANN is trained, the inversion is performed almost instantaneously. Therefore, it can also be used to obtain better initial guesses in order to accelerate traditional inversion methods.\\
It is worth mentioning that the proposed method is not intended to be a general--purpose invertion method, but an specialized one focused in the particular problem of evaluating the calcaneus.

\section*{Acknowledgment}
The authors would like to thank to Dra. Guadalupe Irastorza from Centro Diagnóstico MON, La Plata, Argentina, for the CT images.\\
This work was supported by a grant from the ``Agencia Nacional de
Promoción Científica y Tecnológica de Argentina'' (Ref. PICT-
2016–2303).


\end{document}